\documentclass[12pt,aps,showpacs,floatfix,onecolumn,tightenlines]{revtex4}
\usepackage{epsfig}
\usepackage{psfig}
\usepackage{graphicx}
\usepackage{graphics}
\usepackage{xspace}
\usepackage{amssymb}
\usepackage{amsmath}
\usepackage{latexsym}
\usepackage{natbib}
\usepackage{mathrsfs}
\newcommand{\inieq}{\begin{eqnarray}}            %inizio formula numerata
\newcommand{\fineq}{\end{eqnarray}}            %fine formula numerata 
\newcommand{\diff}{{\rm\,d}}                    %simbolo di derivata totale
\newcommand{\bint}{\mskip .5mu \int \mskip-18mu} %sovrappone il simbolo di
\def\p{\mbox{\boldmath $p$}}

\def\q{\mbox{\boldmath $q$}}

\def\k{\mbox{\boldmath $k$}}

\def\mcg{\mbox{$\mathcal{G}$}}
\def\mcv{\mbox{$\mathcal{V}$}}
\begin{document}
%\begin{frontmatter}
\title{Relativistic Green's function approach to parity-violating
 quasielastic electron scattering} 

\author{Andrea Meucci} 
\author{Carlotta Giusti}
\author{Franco Davide Pacati }
\affiliation{Dipartimento di Fisica Nucleare e Teorica, 
Universit\`{a} degli Studi di Pavia and \\ 
Istituto Nazionale di Fisica Nucleare, 
Sezione di Pavia, I-27100 Pavia, Italy}

%\date{\today}

\begin{abstract}
A relativistic Green's function approach to parity-violating quasielastic
electron scattering is  presented. The components of the hadron tensor are
expressed in terms of the single particle Green's function, which is expanded in
terms of the eigenfunctions of the non-Hermitian optical potential, in order to
account for final state interactions without any loss of flux. Results for
$^{12}$C, $^{16}$O, and $^{40}$Ca are presented and discussed.
The effect of the strange quark contribution to the nuclear current is 
investigated.
\end{abstract}
%\begin{keyword}
%Parity-violating electron scattering \sep
%Quasielastic inclusive electron scattering \sep Relativistic models
%\sep Many-body theory

%\end{keyword} 

\maketitle

%\end{frontmatter}

\section{Introduction}

The study of the nucleon with neutral weak probes has recently gained a wide 
interest in
order to investigate the contribution of the sea quarks to ground state nucleon 
properties, such as spin, charge and magnetic 
moment \cite{kaplan,beck1,beck2}. Besides the
measurements of neutrino-nucleus scattering, experiments of parity-violating (PV)
electron scattering, combined with existing data of nucleon
electromagnetic form factors, may allow to determine possible strange quark
contribution to the spin structure of the 
proton \cite{fein,wale,mck,beck,nap}. First measurements
of PV asymmetry in elastic electron scattering  have been carried
out in recent years. The SAMPLE experiment \cite{mul} at the MIT-Bates 
Laboratory and the HAPPEX collaboration \cite{aniol} at Jefferson Laboratory
(JLab) investigated such asymmetry at $Q^2$ = 0.1 (GeV/c)$^2$ and backward
direction
and 0.5 (GeV/c)$^2$ and forward direction, respectively. The first results 
seemed to indicate a relatively small strangeness contribution to the proton 
magnetic moment \cite{spa,has} and that
the strange form factors must rapidly fall off at large $Q^2$, 
if the strangeness radius is large \cite{aniol}. 
The HAPPEX2 experiment \cite{hap2} at JLab aims at exploring this possibility 
through an improved measurement at smaller $Q^2$. The G0 experiment \cite{g0} at
JLab plans to measure the scattering of electrons by protons both at backward
and forward angles and over the range $0.1 \leq Q^2 \leq 1$ (GeV/c)$^2$ in order
to investigate the strangeness contribution.
The SAMPLE collaboration has recently reported \cite{spa2} a new determination 
of the strange quark contribution to the proton magnetic form factor using a 
revised analysis of data in combination with the axial form factor of the proton
\cite{ito}.  
Another measurement of parity violating asymmetry is going on at Mainz 
Microtron \cite{a4} in order to
determine the combination of strange Dirac and Pauli form factors at
$Q^2$ = 0.225 (GeV/c)$^2$ with great accuracy.
New experiments at JLab \cite{he4p} plan to measure the parity violating
asymmetry using $^4$He and $^{208}$Pb as target nuclei. A recent review 
of the present situation with also a discussion of the theoretical perspectives
of this topic can be found in Ref. \cite{ram}.

In addition to elastic electron scattering, the PV asymmetry can be analyzed 
in inelastic scattering of polarized electrons on nuclei. Besides the inelastic
excitations of discrete states in nuclei \cite{mintz}, the quasielastic (QE) 
electron scattering is the most interesting case. 
In this way, it is possible to understand the role of
the various single-nucleon form factors and, by changing the kinematics and the
target nucleus, to alter the sensitivity to the various responses. However,
nuclear structure effects have to be clearly understood since PV quasielastic
electron scattering introduces new complications concerning the nuclear
responses to neutral current probes. 
General review papers about probes of the hadronic weak neutral current can be 
found in Refs. \cite{Walecka,peccei,musolf,amr,alb,kolbe03}.

A relativistic mean field model of PV observables and strange-quark contribution 
was discussed in Ref. \cite{horo}.
The relativistic Fermi gas (RFG) model was applied to investigate the 
sensitivity to nucleon 
form factors of parity-conserving (PC) and PV responses in QE scattering from $^{12}$C  
in Refs. \cite{alb1,donnelly}, where also strangeness contribution was considered.
Different and more complicated models, including correlations and meson-exchange
currents were later considered in Refs. \cite{alb2,barbaro}. A continuum shell
model description was proposed in Ref. \cite{amaro} and applied to
different closed shell nuclei. 

The effect of final state interactions (FSI) has been stressed to significantly 
contribute to the PC inclusive responses.
Namely, it is essential to explain the exclusive one-nucleon knockout, 
which gives the dominant contribution to the inclusive process in the QE 
region. It is usually described by an
optical potential, whose real component is fitted to elastic proton-nucleus
scattering, while the imaginary part takes into
account the absorption in the final state. 
The reaction channels are thus globally described by
a loss of flux produced by the imaginary part of the complex potential.
This model has been applied with great success to exclusive QE electron 
scattering \cite{book}, where it is able to explain the experimental cross
sections of one-nucleon knockout reactions in a range of nuclei from $^{12}$C to 
$^{208}$Pb. In an inclusive process, however, the flux must be conserved.
This may be obtained by dropping the imaginary part of the optical 
potential and neglecting absorption. However, this procedure conserves the flux but 
it is not consistent with 
the exclusive reaction, which can only be described with a careful treatment 
of the optical potential, including both real and imaginary parts \cite{book}.

We apply a Green's function approach where the conservation of 
flux is preserved and FSI are treated in the inclusive reaction consistently 
with the exclusive one. This method was discussed in a 
nonrelativistic \cite{capuzzi} and in a relativistic framework for 
the case of inclusive PC electron~\cite{ee} and charged-current $\nu$-nucleus
\cite{cc} scattering and 
it is here applied, in a 
relativistic framework, to PV electron scattering. In this approach the 
components of the nuclear response are written in terms 
of the single-particle optical-model Green's function. This result can be  
derived with arguments based on the multiple scattering 
theory~\cite{hori}, on the Feshbach projection 
operator formalism~\cite{capuzzi,chinn,bouch,capma}, and on the mass-operator
properties \cite{kad}. Then, the spectral representation of the 
single-particle Green's function, 
based on a biorthogonal expansion in terms of the eigenfunctions of the 
non-Hermitian optical potential and of its Hermitian conjugate is used to 
perform explicit calculations and to treat FSI consistently in the inclusive 
and in the exclusive reactions. Important 
and peculiar effects are given in the inclusive ($e,e'$) reaction by the 
imaginary part of the optical potential, which is responsible for the  
redistribution of the strength among different channels. 
 
In Sec. \ref{sec.cross} the general formalism of the PV electron 
scattering is given. In Sec. \ref{sec.green}, the Green's function 
approach is briefly reviewed. In Sec. \ref{result}, the results obtained 
on $^{12}$C, $^{16}$O, and $^{40}$Ca target nuclei are presented and discussed. 
Some conclusions are drawn in Sec. \ref{conc}.

\section{Nuclear responses and asymmetry}
\label{sec.cross}

A polarized electron, with 
four-momentum $k^{\mu}_i = (\varepsilon_i,\k_i)$ and longitudinal polarization
$\lambda$, is scattered through an angle 
$\vartheta$ to the final four-momentum $k^{\mu} = (\varepsilon,\k)$ 
via the exchange of a photon or a $Z^0$ with the target nucleus with a 
four-momentum transfer $q^{\mu} = k^\mu_i - k^\mu = (\omega,\q)$. 
The invariant 
amplitude of the process is given to lowest order by the sum of the one-photon
and the one-$Z^0$ boson exchange term. The first term is parity-conserving
whereas the second one has a parity-violating contribution.
The differential cross section is proportional to
\inieq
\diff\sigma \propto \ \mid \mathcal{M}^{\gamma} + \mathcal{M}^{Z} \mid ^2 \simeq 
\ \mid \mathcal{M}^{\gamma} \mid ^2 +\ 
\left(\mathcal{M}^{\gamma}\right)^{\star}\mathcal{M}^{Z} +
(\mathcal{M}^{Z})^{\star}\mathcal{M}^{\gamma}  \ ,
\label{eq.amp}\fineq
where the electromagnetic-weak interference term contains the leading order
PV contribution, while the very small purely weak term 
$\mid \mathcal{M}^{Z} \mid ^2$ can be safely 
neglected.
Eq. \ref{eq.amp} can be rearranged to make explicit the contraction 
between the lepton tensor and the hadron tensor, i.e.,
\inieq
\diff\sigma_{\lambda} \propto L_S^{\mu\nu}W^{\mathrm{S}}_{\mu\nu} + 
\lambda A_0 \left[ g_{\mathrm V}L_A^{\mu\nu} W^{\mathrm{I}}_{\mu\nu}(A)
+ g_{\mathrm A}L_S^{\mu\nu}W^{\mathrm{I}}_{\mu\nu}(V) \right] \ , \label{eq.cs}
\fineq
where the symmetrical and antisymmetrical components, $L_S^{\mu\nu}$ 
and $L_A^{\mu\nu}$, of the lepton tensor are
defined as in Refs. \cite{book,cc,nc}. 
$W^{\mathrm{S}}_{\mu\nu}$ is the symmetrical and unpolarized component of 
the hadron tensor, $W^{\mathrm{I}}_{\mu\nu}(V)$ and $W^{\mathrm{I}}_{\mu\nu}(A)$
are the symmetrical and antisymmetrical polarized components of the hadron
tensor, dependent on vector and axial weak currents, respectively.
The scale factor $A_0$ is defined as
\inieq
A_0 = \frac{G\ Q^2}{2\sqrt{2}\pi\alpha} \simeq 1.798804 \times 10^{-4}
\frac{Q^2}{(\mathrm{GeV/c})^2} \ , \label{eq.a0}
\fineq
where $G\simeq 1.16639 \times 10 ^{-11}$ MeV$^{-2}$ is the Fermi constant, $Q^2
= \mid\q\mid^2 - \omega^2$, $\alpha$ is the fine structure constant and
the couplings $g_{\mathrm A} = -1/2$ and $g_{\mathrm V} = -1/2 + 2
\sin^2{\vartheta_{\mathrm W}} \simeq -0.03714$, where $\vartheta_{\mathrm W}$ is
the Weinberg angle ($\sin^2{\vartheta_{\mathrm W}} \simeq 0.23143$).

The components of the hadron tensor 
are given by suitable bilinear products of the transition matrix
elements of the nuclear current operator $J^{\mu}$ between
the initial state $\mid\Psi_0\rangle$ of the nucleus, of energy $E_0$, and the 
final states $\mid \Psi_{\textrm {f}} \rangle$, of energy $E_{\textrm {f}}$, 
both eigenstates of the $(A+1)$-body Hamiltonian $H$, as
\begin{eqnarray}
& & W^{\mu\nu}(\omega,q) = 
 \bint\sum_{\textrm {f}}  \langle 
\Psi_{\textrm {f}}\mid J^{\mu}(\q) \mid \Psi_0\rangle 
\langle 
\Psi_0\mid J^{\nu\dagger}(\q) \mid \Psi_{\textrm {f}}\rangle 
\ \delta (E_0 +\omega - E_{\textrm {f}}).
\label{eq.ha1}
\end{eqnarray}
where the sum runs over all the states of the residual nucleus.
The single-particle electromagnetic part of the current is 
\begin{eqnarray}
j_{\mathrm{em}}^{\mu} =  F_1(Q^2) \gamma ^{\mu} + 
             i\frac {\kappa}{2M} F_2(Q^2)\sigma^{\mu\nu}q_{\nu}\ .
	     \end{eqnarray}
The single-particle current operator related to the weak neutral 
current is  
\begin{eqnarray}
  j_{\mathrm{nc}}^{\mu} =  F_1^{\textrm V}(Q^2) \gamma ^{\mu} + 
             i\frac {\kappa}{2M} F_2^{\textrm V}(Q^2)\sigma^{\mu\nu}q_{\nu}	  
	     -G_{\textrm A}(Q^2)\gamma ^{\mu}\gamma ^{5} \ ,
	     \label{eq.nc}
\end{eqnarray}
where $\kappa$ is the anomalous part of 
the magnetic moment and
$\sigma^{\mu\nu}=\left(i/2\right)\left[\gamma^{\mu},\gamma^{\nu}\right]$, 
$F_1^{\textrm V}$ and $F_2^{\textrm V}$ are the isovector Dirac and Pauli 
nucleon form factors, and 
$G_{\textrm A}$ is the axial form factor.
The vector form factors $F_i^{\mathrm V}$ can be expressed in terms of the 
corresponding electromagnetic form factors for protons $(F_i^{\mathrm p})$ and 
neutrons $(F_i^{\mathrm n})$, plus a possible isoscalar strange-quark 
contribution $(F_i^{\mathrm s})$, i.e.,
\begin{eqnarray}
F_i^{\mathrm V} &=& \left(\frac{1}{2} - 2\sin^2{\theta_{\mathrm W}}\right)
 F_i^{\mathrm p} -\frac{1}{2} F_i^{\mathrm n} - \frac{1}{2} F_i^{\mathrm s} \ 
 \ \ \ \ ({\mathrm {proton\ knockout}}) \ , \nonumber \\
 F_i^{\mathrm V} &=& \left(\frac{1}{2} - 2\sin^2{\theta_{\mathrm W}}\right)
 F_i^{\mathrm n} -\frac{1}{2} F_i^{\mathrm p} - \frac{1}{2} F_i^{\mathrm s} \ 
 \ \ \ \ ({\mathrm {neutron\ knockout}}) \ , 
\end{eqnarray}
In the calculations the electromagnetic nucleon form factors are taken from 
Ref. \cite{bba}. 
The strange vector form factors are taken as \cite{alb}
\begin{eqnarray}
F_1^{\mathrm s}(Q^2) =  \frac {(\rho^{\mathrm s} + 
\mu^{\mathrm s}) \tau}{(1+\tau) (1+Q^2/M_{\mathrm V}^2)^2} , \ 
F_2^{\mathrm s}(Q^2) =  \frac {\left(\mu^{\mathrm s}-\tau \rho^{\mathrm s}  
\right)}{(1+\tau) (1+Q^2/M_{\mathrm V}^2)^2} ,
\label{eq.sform}
\end{eqnarray}
where $\tau = Q^2/(4M_{\mathrm p}^2)$ and $M_{\mathrm V}$ =
0.843 GeV. The quantities $\mu_{\mathrm s}$ and $\rho_{\mathrm s}$ are related
to the strange magnetic moment and radius of the nucleus. 

The axial form factor is expressed as \cite{mmd}
\begin{eqnarray}
G_{\mathrm A} &=& \frac{1}{2} \left( g_{\mathrm A} - g^{\mathrm s}_{\mathrm
A}\right) G\ \ \ \ ({\mathrm {proton\ knockout}}) \ , \nonumber \\
G_{\mathrm A} &=& -\frac{1}{2} \left( g_{\mathrm A} + g^{\mathrm s}_{\mathrm
A}\right) G\ \ \ \ ({\mathrm {neutron\ knockout}}) \ ,
\end{eqnarray}
where $g_{\mathrm A} \simeq 1.26$, 
$g^{\mathrm s}_{\mathrm A}$ describes possible strange-quark contributions, and 
\begin{equation}
G = (1+Q^2/M_{\mathrm A}^2)^{-2}. 
\end{equation}
The axial mass has been taken 
from Ref. \cite{bernard} as $M_{\mathrm A}$ = (1.026$\pm$0.021) GeV, which is
the weighed average of the values obtained from (quasi)elastic neutrino and
antineutrino scattering experiments.

One can derive from Eq. \ref{eq.cs} the expression for the inclusive 
differential cross section with 
respect to the energy and scattering angle of the final electron. The
parity-conserving inclusive cross section, for unpolarized electron and 
considering only the dominant electromagnetic term of the hadron tensor, is 
\inieq
\left(\frac{\diff \sigma}{\diff \varepsilon \diff \Omega}\right)^{\mathrm{PC}} =
\sigma_{\mathrm{M}} 
\left[ v^{}_{\mathrm{L}}R_{\mathrm{L}} + v^{}_{\mathrm{T}}R_{\mathrm{T}}\right] 
\ , \label{eq.cspc}\fineq
where $\sigma_{\mathrm{M}}$ is the Mott cross section \cite{book}. 
%and $K$ is a kinematical factor \cite{book}. 

The difference of the polarized cross sections gives the parity-violating 
contribution, which is obtained from the interference hadron tensor, i.e.,
\inieq
\left(\frac{\diff \sigma}{\diff \varepsilon \diff \Omega}\right)^{\mathrm{PV}} 
&=&
\frac {1} {2} \left(\frac{\diff \sigma_+}{\diff \varepsilon \diff \Omega} - 
\frac{\diff \sigma_-}{\diff \varepsilon \diff \Omega} \right) \nonumber \\ &=& 
\sigma_{\mathrm{M}}A_0
\Big[ v^{}_{\mathrm{L}}R_{\mathrm{L}}^{\mathrm{AV}} + v^{}_{\mathrm{T}}
R_{\mathrm{T}}^{\mathrm{AV}} 
+ v_{\mathrm{T}}'R_{\mathrm{T}}^{\mathrm{VA}} 
\Big]  ,
\label{eq.cspv}\fineq
where $A_0$ is defined in Eq. \ref{eq.a0}.
The helicity asymmetry can be written as the ratio between the
PV and the PC cross section
\inieq
A =  A_0 \frac {v^{}_{\mathrm{L}}R_{\mathrm{L}}^{\mathrm{AV}} +
 v^{}_{\mathrm{T}}
R_{\mathrm{T}}^{\mathrm{AV}} + v_{\mathrm{T}}'R_{\mathrm{T}}^{\mathrm{VA}}} 
{v^{}_{\mathrm{L}}R_{\mathrm{L}} + v^{}_{\mathrm{T}}R_{\mathrm{T}}}\ .
\label{eq.A}\fineq
The coefficients $v$ are 
\begin{eqnarray}
v^{}_{\mathrm{L}} = \left(\frac {Q^2} {|\q|^2}\right)^2 \ , \ 
v^{}_{\mathrm{T}}=\tan^2\frac {\vartheta}{2} + \frac{Q^2} {2|\q|^2} \ , \ 
v'_{\mathrm{T}}= \tan \frac {\vartheta}{2} \left[ \tan^2\frac {\vartheta}{2} +
 \frac{Q^2} {|\q|^2} \right]^{\frac {1} {2}}   . 
\label{eq.lepton}
\end{eqnarray}
The response functions $R$ are given in terms of the components of the 
hadron tensor as
\begin{eqnarray} 
R_{\mathrm{L}} &=&  W_{00}^{\mathrm{em}}\ , \ R_{\mathrm{T}}=
\left(W_{xx}^{\mathrm{em}} + W_{yy}^{\mathrm{em}} \right)\ , 
\nonumber \\
R_{\mathrm{L}}^{\mathrm{AV}} &=& g_{\mathrm A}W_{00}^{\mathrm{I}}\ , \ 
R_{\mathrm{T}}^{\mathrm{AV}} = g_{\mathrm A}\left(W_{xx}^{\mathrm{I}}
+ W_{yy}^{\mathrm{I}}\right)\ , \nonumber \\
R_{\mathrm{T}}^{\mathrm{VA}} &=& i g_{\mathrm V}\left(W_{xy}^{\mathrm{I}} - 
W_{yx}^{\mathrm{I}}\right)\ ,
\label{eq.rf}
\end{eqnarray}
where the superscript AV denotes interference of axial-vector leptonic current
with vector hadronic current (the reverse for VA).

\section{The relativistic Green's function approach}
\label{sec.green}

We apply here to the inclusive PV electron scattering the same relativistic 
approach which was already applied to the inclusive PC electron 
scattering \cite{ee} and to the inclusive 
QE $\nu$($\bar\nu$)-nucleus scattering \cite{cc}. 
Here we recall only the most important features
of the model. More details can be found in Refs.~\cite{book,capuzzi,ee} 

For the inclusive process the components of the hadron tensor can be expressed 
as 
\inieq
 W^{\mu\nu}(\omega,q)  
=\langle \Psi_0 \mid J^{\nu\dagger}(\q) \delta(E_{\textrm {f}}-H) 
J^{\mu}(\q) \mid \Psi_0 \rangle \ . 
\label{eq.hadrontensor}
\fineq
Using the equivalence
\inieq
\delta(E-H) = \frac {1} {2 \pi i} [G^{\dagger}(E) - G(E)] \ ,
\label{eq.delta}
\fineq
in terms of the Green's operators
\inieq
  G^{\dagger}(E) = \frac{1} {E - H - i\eta} \ , \, \,
  G(E) = \frac{1} {E - H + i\eta} \ ,  \label{eq.green}
\fineq  
related to the nuclear Hamiltonian $H$, we have 
\inieq
 \omega^{\mu\mu} =  W^{\mu\mu}(\omega,q)  
= -\frac{1}{\pi} \textrm{Im} \langle \Psi_0 \mid J^{\mu\dagger}(\q) 
G(E_{\textrm {f}}) J^{\mu}(\q) \mid \Psi_0 \rangle \ , 
\label{eq.ht1}
\fineq
and
\inieq
 \omega^{\mu\nu} &=&  W^{\mu\nu}(\omega,q) \pm W^{\nu\mu}(\omega,q) 
 \nonumber \\ &=&
-\frac{1}{\pi} \textrm{Im} \langle \Psi_0 \mid J^{\nu\dagger}(\q) 
G(E_{\textrm {f}}) J^{\mu}(\q) \pm J^{\mu\dagger}(\q)G(E_{\textrm {f}}) 
J^{\nu}(\q) \mid \Psi_0 \rangle \ , 
\label{eq.ht1bis}
\fineq
for $\mu \not= \nu$, where the upper (lower) sign refers to the symmetrical 
(antisymmetrical) components of the hadron tensor.

It was shown in Refs. \cite{ee,cc} that the nuclear response in 
Eq. \ref {eq.hadrontensor} can be written in terms of 
the single particle Green's function, $\mcg(E)$, whose self-energy is the 
Feshbach's optical potential. 
A biorthogonal expansion of the full particle-hole Green's operator is then 
performed in terms of the 
eigenfunctions of the non-Hermitian optical potential
$\mcv$ and of its Hermitian conjugate $\mathcal{V}^{\dagger}$,
\inieq
\left[ {\mathcal{E}} - T - {\mathcal{V}}^{\dagger} (E) \right] \mid
{\chi}_{\mathcal{E}}^{(-)}(E)\rangle = 0\ , \ \    %\nonumber \\
\left[ \mathcal{E} - T - {\mathcal{V}}(E) \right] \mid \tilde
{\chi}_{\mathcal{E}}^{(-)}(E)\rangle = 0\ ,
 \label{eq.op}
\fineq
where $E$ and ${\mathcal{E}}$ are not necessarily the same.
The spectral representation of $\mcg(E)$ is
\inieq
\mcg(E) = \int_M^{\infty} \diff \mathcal{E}\mid\tilde
{\chi}_{\mathcal{E}}^{(-)}(E)\rangle 
 \frac{1}{E-\mathcal{E}+i\eta} \langle\chi_{\mathcal{E}}^{(-)}(E)\mid 
\ . \label{eq.sperep}
\fineq
The hadron tensor components can be reduced to a single-particle 
expression and $\omega^{\mu\nu}$ can be written in an expanded form as
\inieq
\omega^{\mu\nu}(\omega , q) = -\frac{1}{\pi} \sum_n  \textrm{Im} \bigg[
 \int_M^{\infty} \diff \mathcal{E} \frac{1}{E_{{\mathrm
{f}}}-\varepsilon_n-\mathcal{E}+i\eta}  
  T_n^{\mu\nu}(\mathcal{E} ,E_{{\mathrm{f}}}-\varepsilon_n) \bigg]
\ , \label{eq.pracw}
\fineq
where $n$ denotes the eigenstate $\mid\ n \rangle$ of the residual Hamiltonian 
of $A$ interacting nucleons related to the discrete eigenvalue $\varepsilon_n$. 
The matrix elements $T^{\mu\nu}$  are defined in terms of the current 
operators. For the
components of $W_{\mathrm{em}}^{\mu\mu}$ we have
\inieq
T_{n,\mathrm{em}}^{\mu\mu}(\mathcal{E} ,E) &=& \lambda_n\langle \varphi_n
\mid j_{\mathrm{em}}^{\mu\dagger}(\q) \sqrt{1-\mcv'(E)}
\mid\tilde{\chi}_{\mathcal{E}}^{(-)}(E)\rangle \nonumber \\
&\times&  \langle\chi_{\mathcal{E}}^{(-)}(E)\mid  \sqrt{1-\mcv'(E)} 
j_{\mathrm{em}}^{\mu}
(\q)\mid \varphi_n \rangle  \ , \label{eq.tpem}
\fineq
for $\mu = 0,x,y$. The components of $W_{\mathrm{I}}^{\mu\mu}$  are
\inieq
T_{n,\mathrm{I}}^{\mu\mu}(\mathcal{E} ,E) &=& \lambda_n[\langle \varphi_n
\mid j_{\mathrm{em}}^{\mu\dagger}(\q) \sqrt{1-\mcv'(E)}
\mid\tilde{\chi}_{\mathcal{E}}^{(-)}(E)\rangle \nonumber \\
&\times&  \langle\chi_{\mathcal{E}}^{(-)}(E)\mid  \sqrt{1-\mcv'(E)} 
j_{\mathrm{nc}}^{\mu}
(\q)\mid \varphi_n \rangle \nonumber \\ &+&
\langle \varphi_n
\mid j_{\mathrm{nc}}^{\mu\dagger}(\q) \sqrt{1-\mcv'(E)}
\mid\tilde{\chi}_{\mathcal{E}}^{(-)}(E)\rangle \nonumber \\
&\times&  \langle\chi_{\mathcal{E}}^{(-)}(E)\mid  \sqrt{1-\mcv'(E)} 
j_{\mathrm{em}}^{\mu}
(\q)\mid \varphi_n \rangle] \ , \label{eq.tpi1}
\fineq
for $\mu = 0,x,y$, and
\inieq
T_{n,\mathrm{I}}^{xy}(\mathcal{E} ,E) &=& \lambda_n i [\langle \varphi_n
\mid j_{\mathrm{em}}^{y\dagger}(\q) \sqrt{1-\mcv'(E)}
\mid\tilde{\chi}_{\mathcal{E}}^{(-)}(E)\rangle \nonumber \\
&\times&  \langle\chi_{\mathcal{E}}^{(-)}(E)\mid  \sqrt{1-\mcv'(E)} 
j_{\mathrm{nc}}^{x} (\q)\mid \varphi_n \rangle \nonumber \\ 
& - & \langle \varphi_n
\mid j_{\mathrm{em}}^{x\dagger}(\q) \sqrt{1-\mcv'(E)}
\mid\tilde{\chi}_{\mathcal{E}}^{(-)}(E)\rangle \nonumber \\
&\times&  \langle\chi_{\mathcal{E}}^{(-)}(E)\mid  \sqrt{1-\mcv'(E)} 
j_{\mathrm{nc}}^{y} (\q)\mid \varphi_n \rangle \nonumber \\
&+&\langle \varphi_n
\mid j_{\mathrm{nc}}^{y\dagger}(\q) \sqrt{1-\mcv'(E)}
\mid\tilde{\chi}_{\mathcal{E}}^{(-)}(E)\rangle \nonumber \\
&\times&  \langle\chi_{\mathcal{E}}^{(-)}(E)\mid  \sqrt{1-\mcv'(E)} 
j_{\mathrm{em}}^{x} (\q)\mid \varphi_n \rangle \nonumber \\ 
& - & \langle \varphi_n
\mid j_{\mathrm{nc}}^{x\dagger}(\q) \sqrt{1-\mcv'(E)}
\mid\tilde{\chi}_{\mathcal{E}}^{(-)}(E)\rangle \nonumber \\
&\times&  \langle\chi_{\mathcal{E}}^{(-)}(E)\mid  \sqrt{1-\mcv'(E)} 
j_{\mathrm{em}}^{y} (\q)\mid \varphi_n \rangle ]\ .
\label{eq.tpi2}
\fineq
The factor 
 $\sqrt{1-\mcv'(E)}$ accounts for
interference effects between different channels and allows the replacement of
the mean field $\mcv$ by the phenomenological optical potential 
$\mcv_{ \mathrm L} $\cite{ee}. $\lambda_n$ is the spectral strength 
\cite{bofficapuzzi} of the hole state
$\mid \varphi_n \rangle$, that is the normalized overlap between $\mid
\Psi_0\rangle$ and $\mid n \rangle$.   
After calculating the limit for $\eta \rightarrow +0$, 
Eq. \ref{eq.pracw} reads
\inieq
\omega^{\mu\nu}(\omega , q) = \sum_n \Bigg[ \textrm{Re} T_n^{\mu\nu}
(E_{\mathrm{f}}-\varepsilon_n, E_{ \mathrm{f}}-\varepsilon_n)  \nonumber \\
- \frac{1}{\pi} \mathcal{P}  \int_M^{\infty} \diff \mathcal{E} 
\frac{1}{E_{\mathrm{f}}-\varepsilon_n-\mathcal{E}} %\nonumber \\
%\times 
\textrm{Im} T_n^{\mu\nu}
(\mathcal{E},E_{\mathrm{f}}-\varepsilon_n) \Bigg] \ , \label{eq.finale}
\fineq
where $\mathcal{P}$ denotes the principal value of the integral. 

Disregarding the square root correction, due to interference effects,
the second matrix element in Eq. \ref{eq.tpem}, with the inclusion of 
$\sqrt{\lambda_n}$, is the transition amplitude for the single-nucleon 
knockout 
from a nucleus in the state $\mid \Psi_0\rangle$ leaving the residual nucleus 
in the state $\mid n \rangle$. The attenuation of its strength,
mathematically due to the imaginary part of the optical potential, is related to 
the
flux lost towards the channels different from $n$. In the inclusive response
this attenuation must be compensated by a corresponding gain, due to the flux
lost, towards the channel $n$, by the other final states asymptotically
originated by the channels different from $n$. 
This compensation is
performed by the first matrix element in the right hand side of 
Eq. \ref{eq.tpem}, where the imaginary part of the potential has the effect of 
increasing the strength. Similar considerations can be made, on the purely 
mathematical ground, for the integral of Eq. \ref{eq.finale}, where the 
amplitudes involved in $T_n^{\mu\nu}$ have no evident physical meaning when 
${\mathcal{E}}\neq E_{\rm{f}}-\varepsilon_n$. 

In an usual shell-model calculation the cross section is obtained from the sum, 
over all the single-particle shell-model states, of the squared absolute value
of the transition matrix elements.
Therefore, in such a calculation the negative imaginary part of the
optical potential produces a loss of flux that is inconsistent with the 
inclusive process.
In the Green's function approach the flux is conserved, as the
components of the hadron tensor are obtained in terms of the product of the two
matrix elements in Eq. \ref{eq.tpem}:  
the loss of flux, produced by the
negative imaginary part of the optical potential in $\chi$, is compensated by 
the gain of flux produced in the first matrix element by the positive 
imaginary part of the Hermitian conjugate optical potential in $\tilde \chi$. 

The cross sections and the response functions are calculated from the 
single-particle expression of the hadron tensor in  Eq. \ref{eq.finale}. 
After the replacement of the mean field $\mcv(E)$ by the empirical optical 
model potential $\mcv_{\textrm {L}}(E)$, the matrix elements of the nuclear 
current operator in Eqs. \ref{eq.tpem}-\ref{eq.tpi2}, which 
represent the 
main ingredients 
of the calculation, are of the same kind as those giving the transition 
amplitudes  of the electron induced nucleon knockout reaction in the 
relativistic distorted wave impulse approximation 
(RDWIA)~\cite{meucci1,meucci2}. 

The relativistic final wave function is written, as in
Refs.~\cite{ee,meucci1,meucci2,meucci3},
in terms of its upper
component following the direct Pauli reduction scheme, i.e.,
\inieq
 \chi_{\mathcal{E}}^{(-)}(E)  =  \left(\begin{array}{c} 
{\displaystyle \Psi_{\textrm {f}+}} \\ 
\frac{\displaystyle 1} {\displaystyle 
M+{\mathcal E}+S^{\dagger}(E)-V^{\dagger}(E)}
{\displaystyle \mbox{\boldmath $\sigma$}\cdot\p
        \Psi_{\textrm {f}+}} \end{array}\right) \ ,
\fineq
where $S(E)$ and $V(E)$ are the scalar and vector 
energy-dependent
components of the relativistic optical potential for a nucleon
with energy $E$ \cite{chc}. 
The upper component, $\Psi_{\textrm {f}+}$, is related to a two-component
spinor, $\Phi_{\textrm{f}}$, which solves a
Schr\"odinger-like equation containing equivalent central and 
spin-orbit potentials, obtained from the relativistic scalar and vector 
potentials \cite{clark,HPa}, i.e.,
\inieq
\Psi_{\textrm {f}+} = \sqrt{D_{{\mathcal E}}^{\dagger}(E)}\ \Phi_{\textrm{f}} \ , 
\quad D_{{\mathcal E}}(E) &=& 1 + \frac{S(E)-V(E)}{M+{\mathcal E}} \ , 
\label{eq.darw}
\fineq
where $D_{{\mathcal E}}(E)$ is the Darwin factor. 

The wave functions $\varphi_n$ are taken as 
the Dirac-Hartree solutions of a relativistic Lagrangian
containing scalar and vector potentials \cite{adfx,lala}.

%%%%%%%%%%%%%%%%%%%%%%%%%%%%%%%%%%%%%%%%%%%%%%%
\section{Results}
\label{result}

The calculations have been performed with the same
bound state wave functions and optical potentials as in
Refs.~\cite{ee,cc,nc,meucci1,meucci2,meucci3,rm}, where the RDWIA was successfully 
applied to 
study $\left(e,e^{\prime}p\right)$, $\left(\gamma,p\right)$, 
$\left(e,e^{\prime}\right)$ and ($\nu-$nucleus) reactions. 

The relativistic bound state wave functions have been obtained as the
Dirac-Hartree solutions of a relativistic Lagrangian containing scalar and vector
potentials deduced in the context of a relativistic mean field theory that
satisfactorily reproduces
single-particle properties of several spherical and deformed 
nuclei~\cite{adfx,lala}. 
The scattering state is calculated by means of
the energy-dependent and A-dependent EDAD1 complex phenomenological optical 
potential of Ref.~\cite{chc}, that is fitted to proton
elastic scattering data on several nuclei in an energy range up to 1040 MeV.

The initial states $\mid \varphi_n \rangle$ are taken as
single-particle one-hole states in the target with a unitary spectral strength. 
The sum runs over all the occupied states.

The results obtained in the Green's function approach are compared with those 
given by different approximations
in order to show up the effect of the optical potential on the inclusive 
responses. 
In the simplest approach the optical potential is neglected, i.e., 
$\mathcal{V} = \mathcal{V}^{\dagger} =0$ in Eq. \ref{eq.op}, and the plane 
wave approximation (PWIA) is assumed for the final state wave functions ${\chi}^{(-)}$
and  $\tilde{\chi}^{(-)}$. In this approximation  FSI
between the outgoing nucleon and the residual nucleus are completely neglected. 
In another approach the integrated contribution of all the single-nucleon
knockout processes is considered. In this case the negative imaginary part of the
optical potential produces a loss of flux that is inconsistent with the 
inclusive process and results in an underestimation of the responses.

\begin{figure}[h]
\begin{center}
\includegraphics[height=14cm, width=12cm]{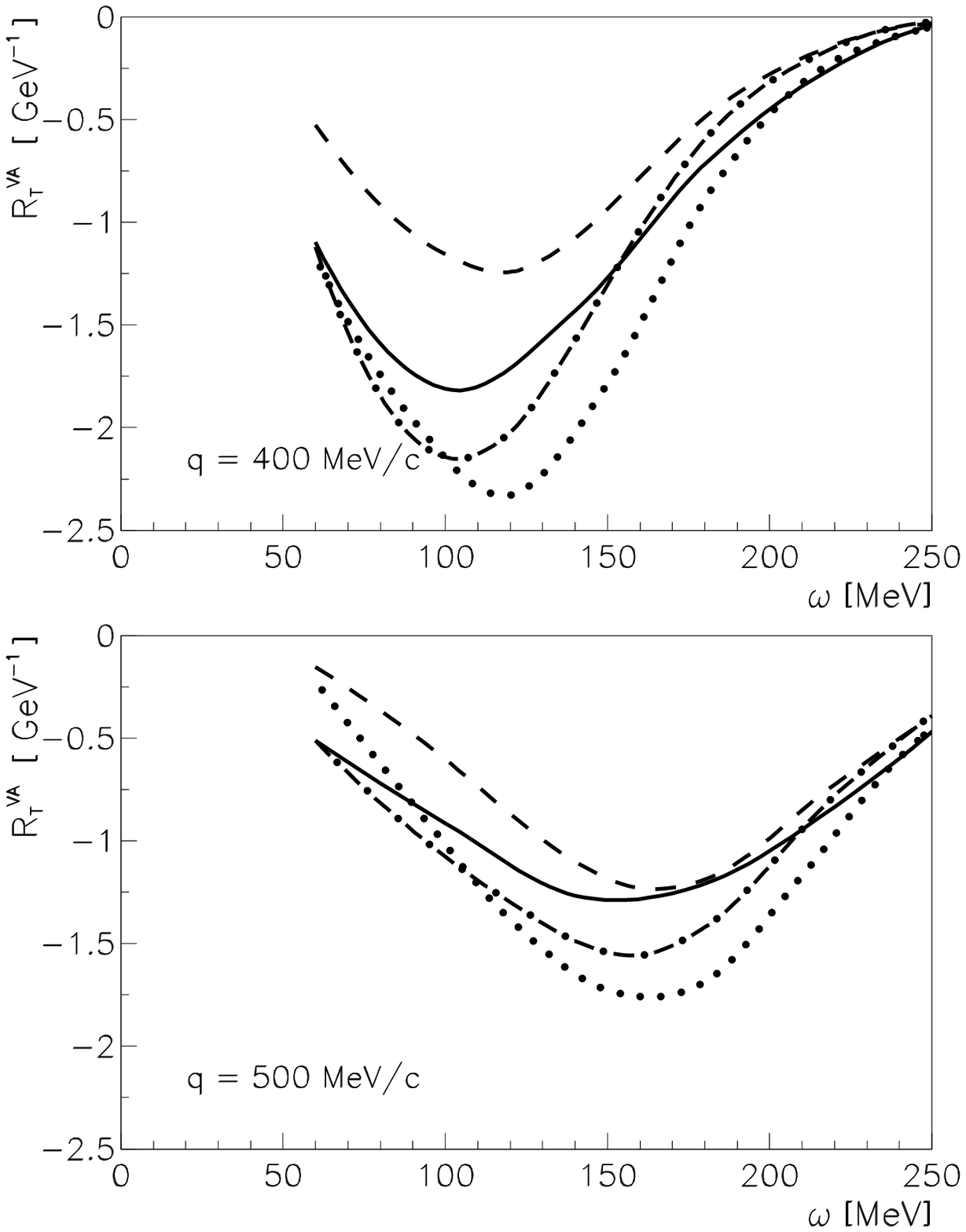} 
\vskip -0.3cm
\caption {The response function $R_{\mathrm{T}}^{\mathrm{VA}}$ of 
the $^{12}$C$(e,e')$ 
reaction for $q$ = 400 and 500 MeV/c.  
Solid lines represent the result of the Green's function approach,
dotted lines give PWIA, dot-dashed lines show the result without the integral 
in Eq. \ref{eq.finale}, and dashed lines the integration of the
exclusive reactions with one-nucleon emission.
 }
\label{f1c}
\end{center}
\end{figure}

\begin{figure}[h]
\begin{center}
\includegraphics[height=14cm, width=12cm]{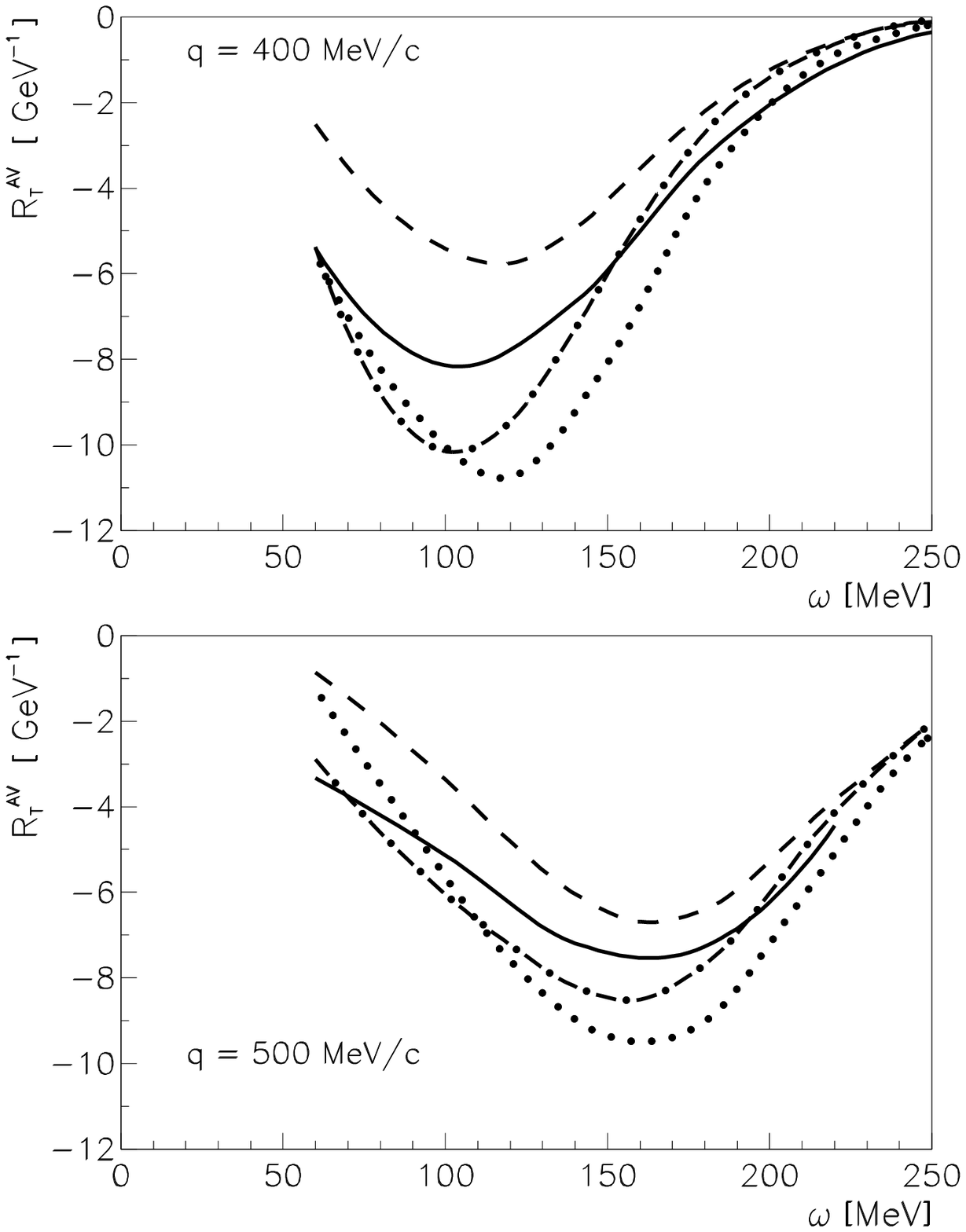} 
\vskip -0.3cm
\caption {The same as in Fig. \ref{f1c}, but for the 
$R_{\mathrm{T}}^{\mathrm{AV}}$response. }
\label{f2c}
\end{center}
\end{figure}

\begin{figure}[h]
\begin{center}
\includegraphics[height=14cm, width=12cm]{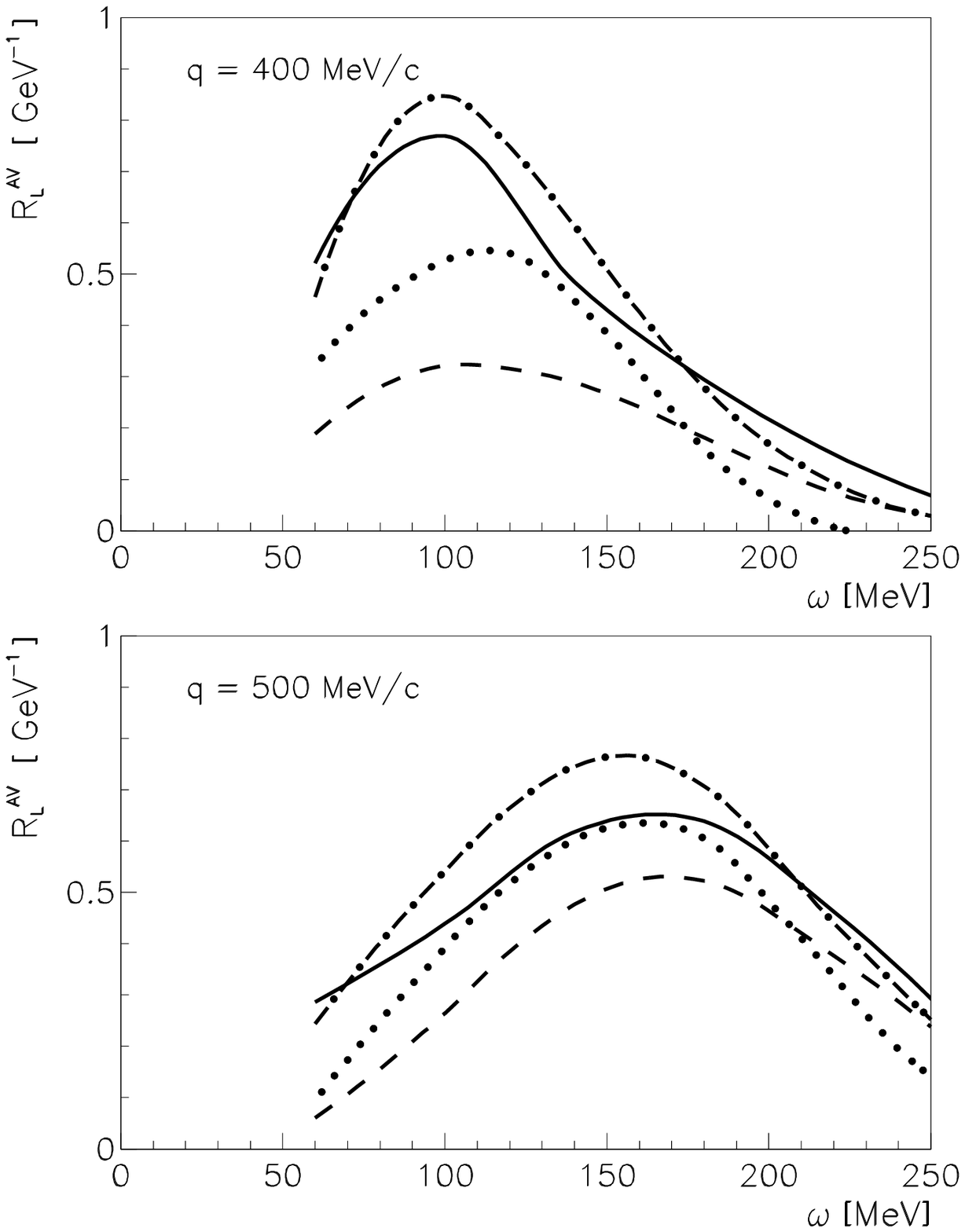} 
\vskip -0.3cm
\caption {The same as in Fig. \ref{f1c}, but for the 
$R_{\mathrm{L}}^{\mathrm{AV}}$response.}
\label{f3c}
\end{center}
\end{figure}

First, we have considered the $^{12}$C$(e,e')$ reaction at
momentum transfer $q$ = 400 and 500 MeV/c. This kinematics corresponds
to that of the experiments performed at Saclay \cite{saclay}. 
In Fig. \ref{f1c} our results for the 
$R_{\mathrm{T}}^{\mathrm{VA}}$ response function are displayed and compared 
with the other approaches, i.e., PWIA and the integration of all the  
single-nucleon knockout channels. The PWIA results are generally
larger than the Green's function ones; moreover, a shift of the position of the
maximum is visible. The contribution of the single-nucleon emission is smaller
than the complete calculation. The difference, that can be attributed to the 
loss of flux produced by the imaginary part of the optical potential, gives an 
idea of the relevance of the inelastic channels.  
In Fig. \ref{f1c} the results obtained with only the
first term of Eq. \ref{eq.finale} are also shown. This term can be
neglected in a nonrelativistic calculation \cite{capuzzi}, where it gives only 
a very small contribution, but must be included in the relativistic
approach \cite{ee}, where it is essential to reproduce the
experimental longitudinal response. In accordance with Ref. \cite{ee}, the 
contribution of the integral in Eq. \ref{eq.finale} gives a 10-15\% 
reduction of the maximum at the momentum transfers considered in Fig. \ref{f1c} 
and becomes less important for larger values of the momentum transfer. 
Similar results are obtained for the $R_{\mathrm{T}}^{\mathrm{AV}}$  and 
$R_{\mathrm{L}}^{\mathrm{AV}}$ responses, as it can be seen in 
Figs. \ref{f2c} and \ref{f3c}. For $R_{\mathrm{L}}^{\mathrm{AV}}$, the PWIA 
results are smaller than the complete ones. This response function, however, 
is only a small fraction of the leading response 
$R_{\mathrm{T}}^{\mathrm{AV}}$ (see also 
Refs. \cite{musolf,donnelly}). It has been argued \cite{barbaro}
that correlations, which are not included in our calculations, can affect this 
particular response at low and moderate momentum
transfer whereas they mildly influence the other responses. 

\begin{figure}[h]
\begin{center}
\includegraphics[height=14cm, width=12cm]{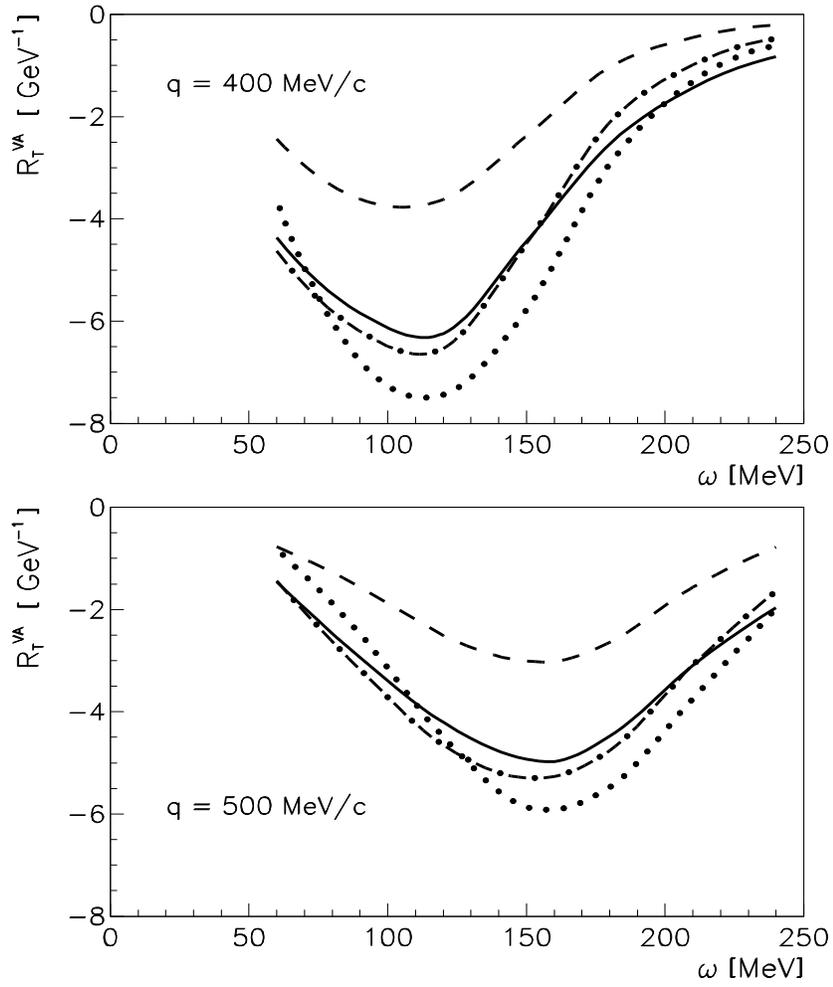} 
\vskip -0.3cm
\caption {The response function $R_{\mathrm{T}}^{\mathrm{VA}}$ of 
the $^{40}$Ca$(e,e')$ 
reaction for $q$ = 400 and 500 MeV/c.
Line convention as in Fig. \ref{f1c}.
 }
\label{f1ca}
\end{center}
\end{figure}

\begin{figure}[h]
\begin{center}
\includegraphics[height=14cm, width=12cm]{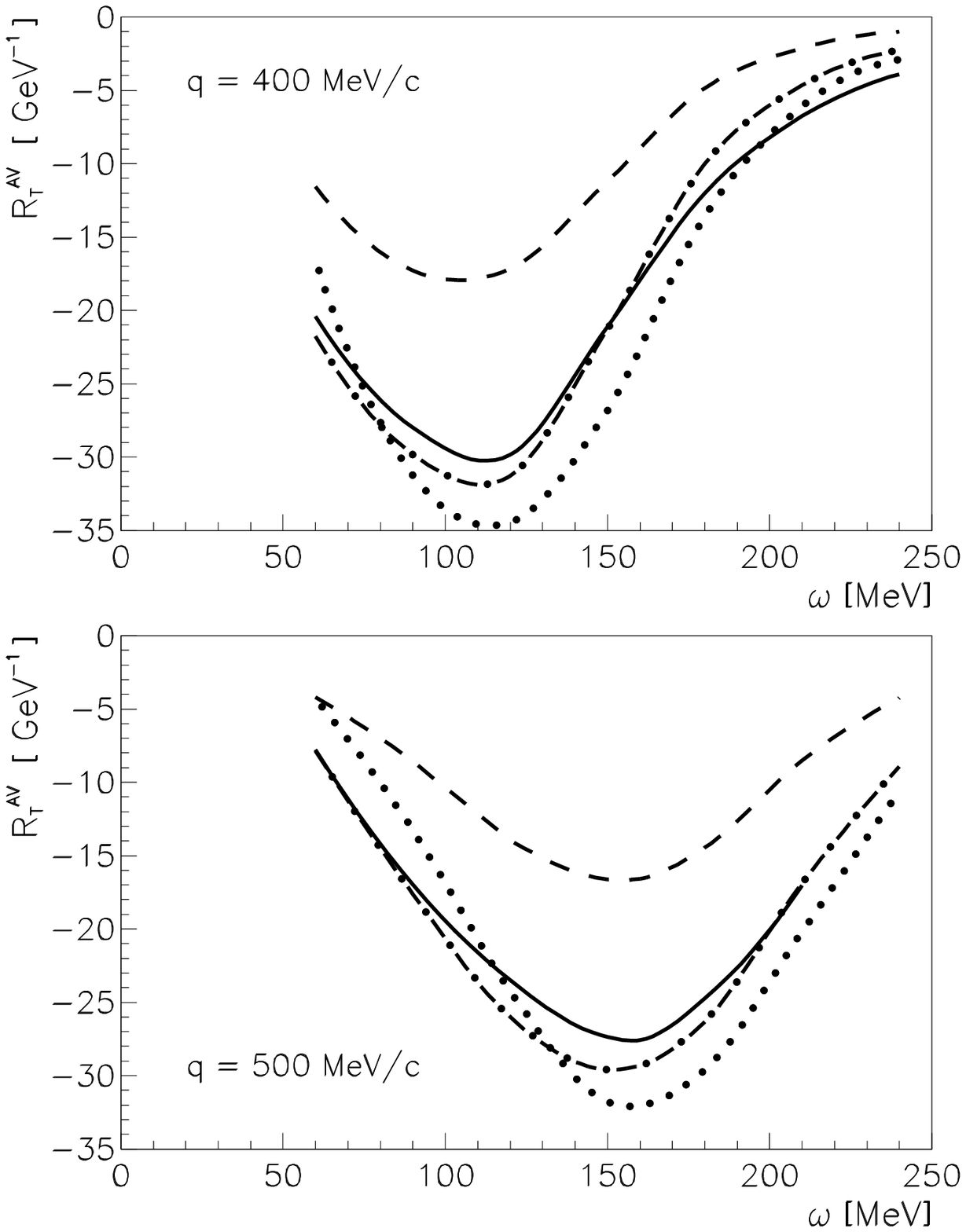} 
\vskip -0.3cm
\caption {The same as in Fig. \ref{f1ca}, but for the 
$R_{\mathrm{T}}^{\mathrm{AV}}$response. }
\label{f2ca}
\end{center}
\end{figure}

\begin{figure}[h]
\begin{center}
\includegraphics[height=14cm, width=12cm]{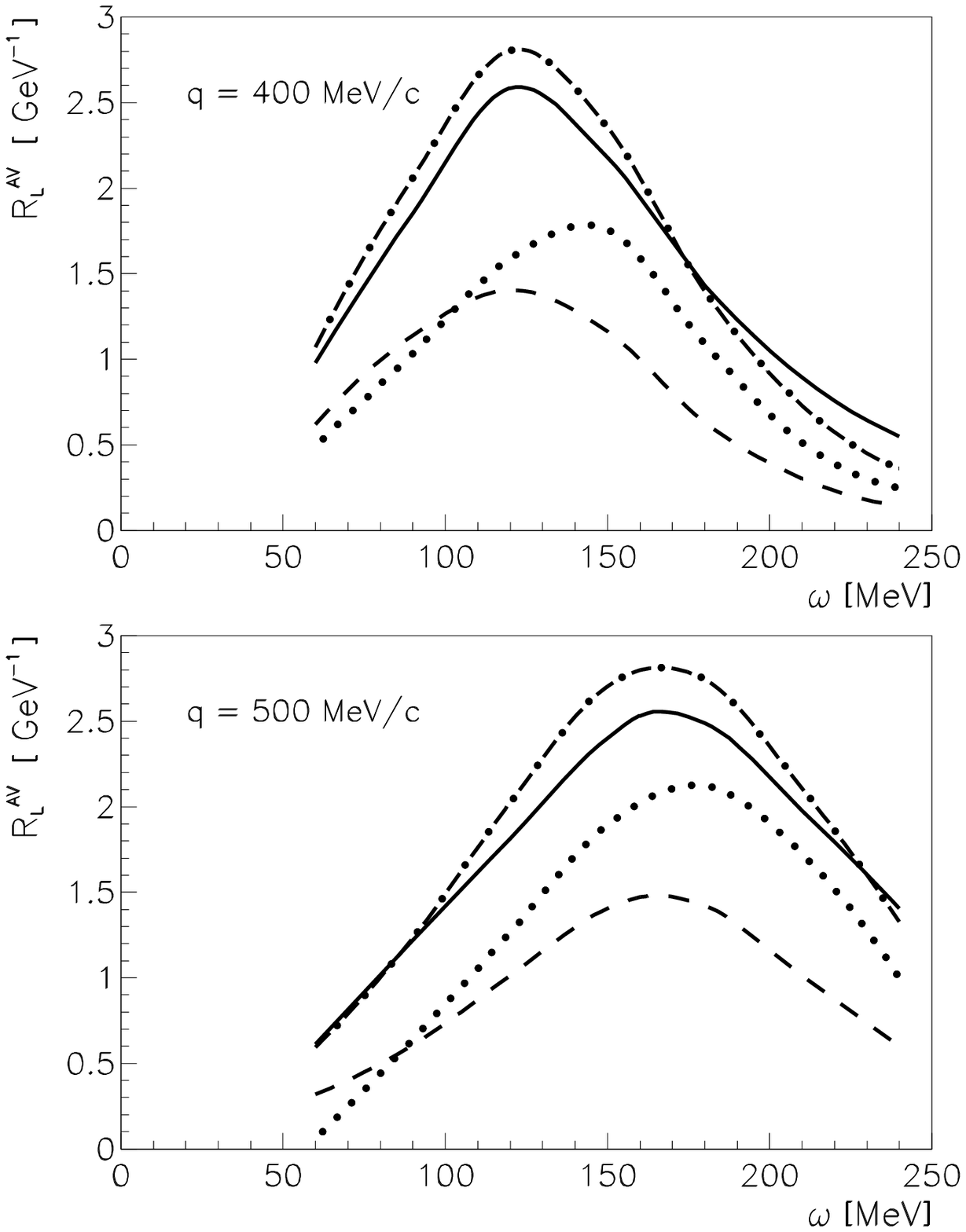} 
\vskip -0.3cm
\caption {The same as in Fig. \ref{f1ca}, but for the 
$R_{\mathrm{L}}^{\mathrm{AV}}$response.}
\label{f3ca}
\end{center}
\end{figure}

In Figs. \ref{f1ca}, \ref{f2ca}, and \ref{f3ca} the same PV
responses are shown
for the $^{40}$Ca$(e,e')$ reaction at $q$ = 400 and 500 MeV/c. The
results are qualitatively similar to those obtained for $^{12}$C. 

\begin{figure}[h]
\begin{center}
\includegraphics[height=14cm, width=12cm]{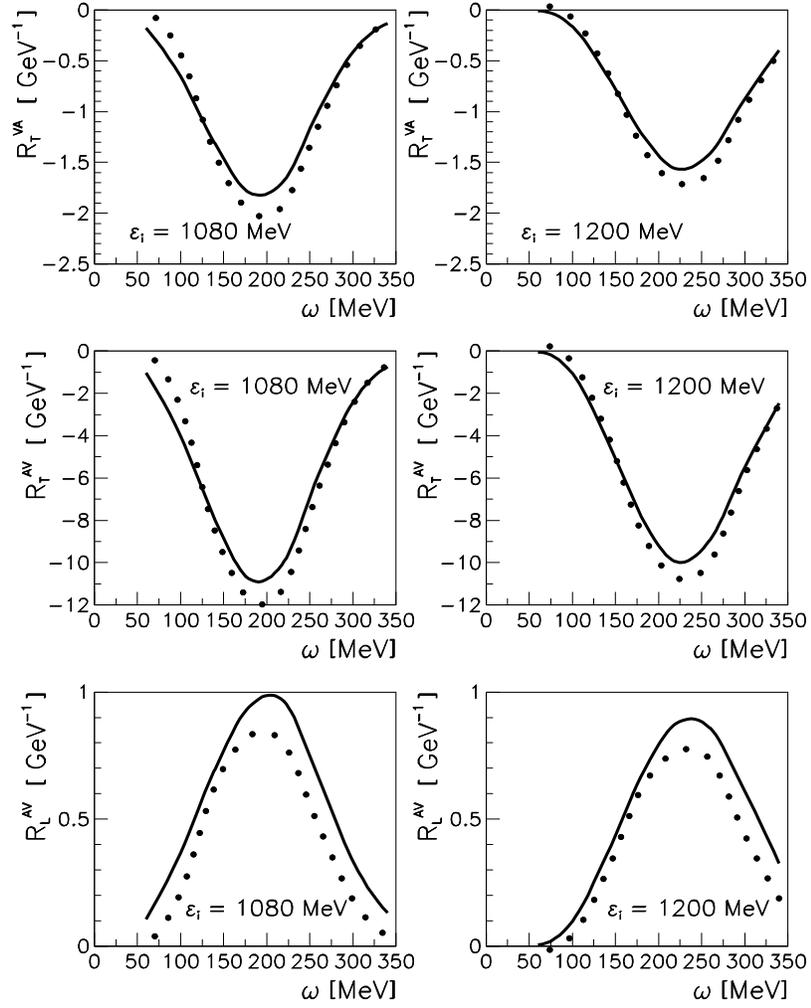} 
\vskip -0.3cm
\caption {The response functions $R_{\mathrm{T}}^{\mathrm{VA}}$, 
$R_{\mathrm{T}}^{\mathrm{AV}}$, and $R_{\mathrm{L}}^{\mathrm{AV}}$ of 
the $^{16}$O$(e,e')$ 
reaction for $\varepsilon_i$ = 1080 (left panels) and 1200 (right panels) MeV and 
$\vartheta = 32^{\mathrm o}$.  
Solid lines represent the result of the Green's function approach and
dotted lines give PWIA.
 }
\label{f1o}
\end{center}
\end{figure}

In Fig. \ref{f1o} the PV responses
are presented for the $^{16}$O$(e,e')$ reaction in a kinematics with beam 
energies
$\varepsilon_i$ = 1080 and 1200 MeV and scattering angle 
$\vartheta$ = 32$^{\mathrm o}$. This choice corresponds to the Frascati 
kinematics \cite{frascati}. The Green's function results are 
compared with the PWIA ones. In
this kinematics at $q$ = 600 MeV/c the contribution of the integral in 
Eq. \ref{eq.finale} is small. 

\begin{figure}[h]
\begin{center}
\includegraphics[height=14cm, width=12cm]{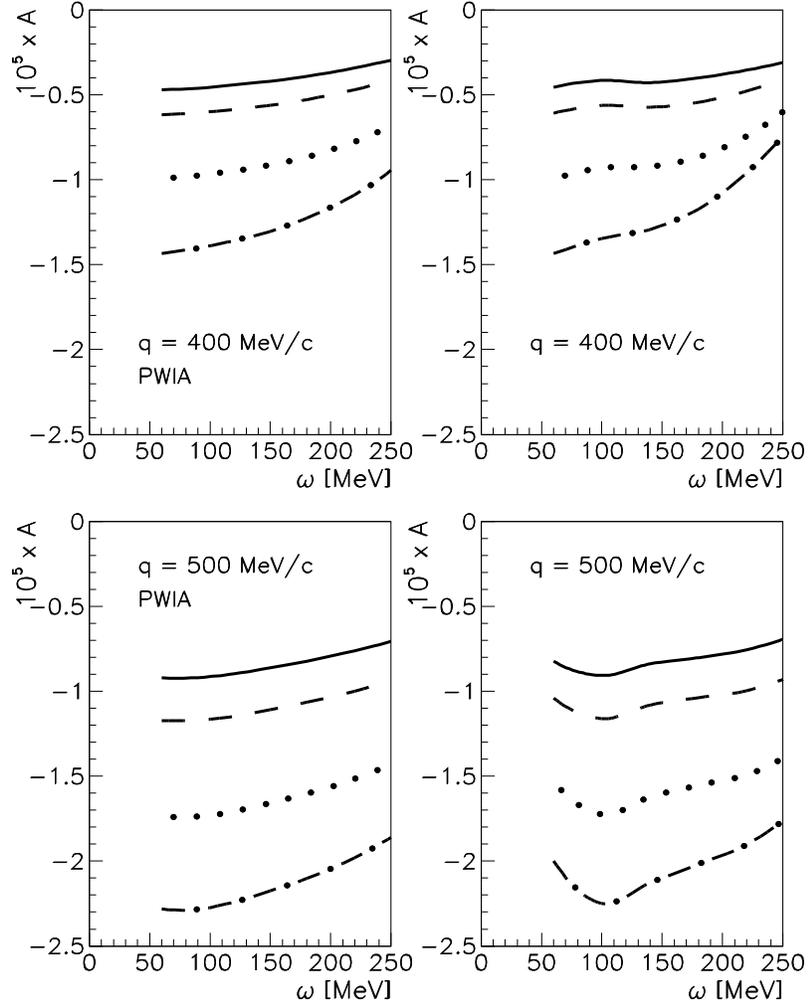} 
\vskip -0.3cm
\caption {PV asymmetry for $^{12}$C 
at $q$ = 400 and 500 MeV/c. Left panels: PWIA results. Right panels: Green's 
function approach. The results are rescaled by the factor
$10^5$. Solid lines represent the results at 
$\vartheta$ = 15$^{\mathrm o}$, dashed lines at 45$^{\mathrm o}$, dotted lines at 
90$^{\mathrm o}$, and dot-dashed lines at 135$^{\mathrm o}$.
}
\label{fa1}
\end{center}
\end{figure}

\begin{figure}[h]
\begin{center}
\includegraphics[height=14cm, width=12cm]{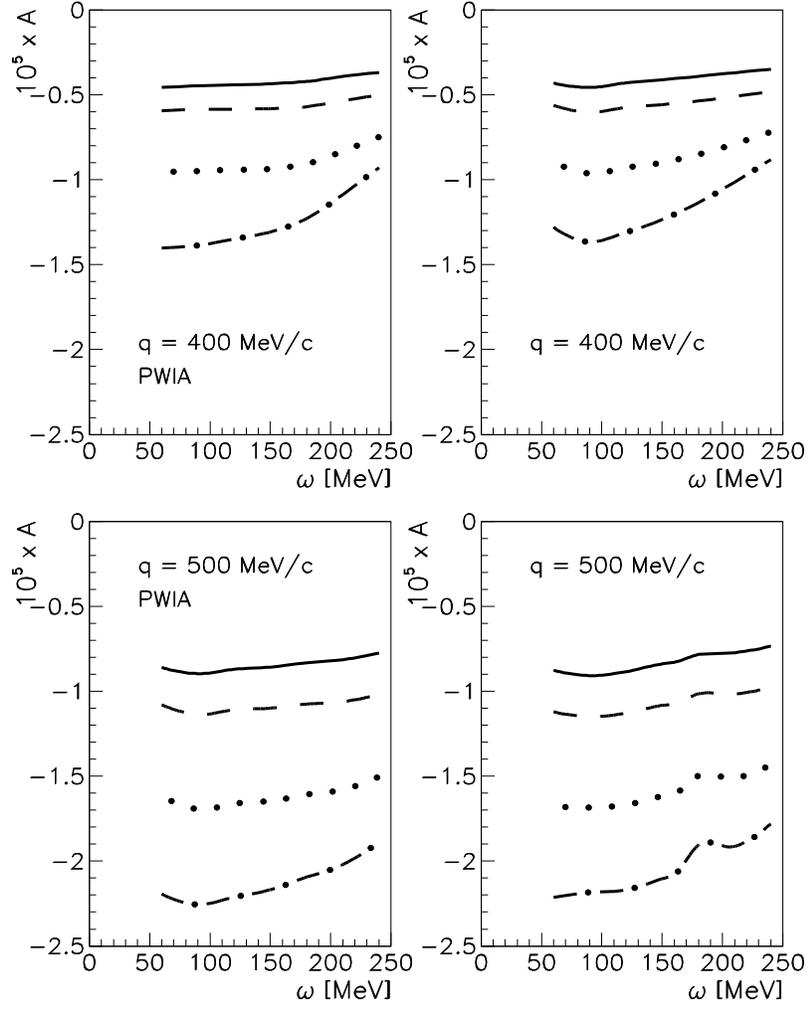} 
\vskip -0.3cm
\caption {The same as in Fig. \ref{fa1}, but for $^{40}$Ca$(e,e')$ 
reaction.
}
\label{fa2}
\end{center}
\end{figure}

In Figs. \ref{fa1} and \ref{fa2} the PV asymmetry of Eq.
\ref{eq.A} for $^{12}$C and $^{40}$Ca at 
$q$ = 400 and 500 MeV/c are displayed at four different values of the
electron scattering angle. Note that the results are rescaled by the factor
$10^5$. The asymmetry
ranges from a few $\times 10 ^{-6}$ at forward angle and low momentum transfer 
to a few $\times 10 ^{-5}$ at backward angle and greater $q$. The results given 
by the Green's function approach are compared with the PWIA ones. Only small 
differences are found: the Green's function results are lower in absolute value 
than the PWIA ones and the shape of the curves is slightly different. 

\begin{figure}[h]
\begin{center}
\includegraphics[height=14cm, width=12cm]{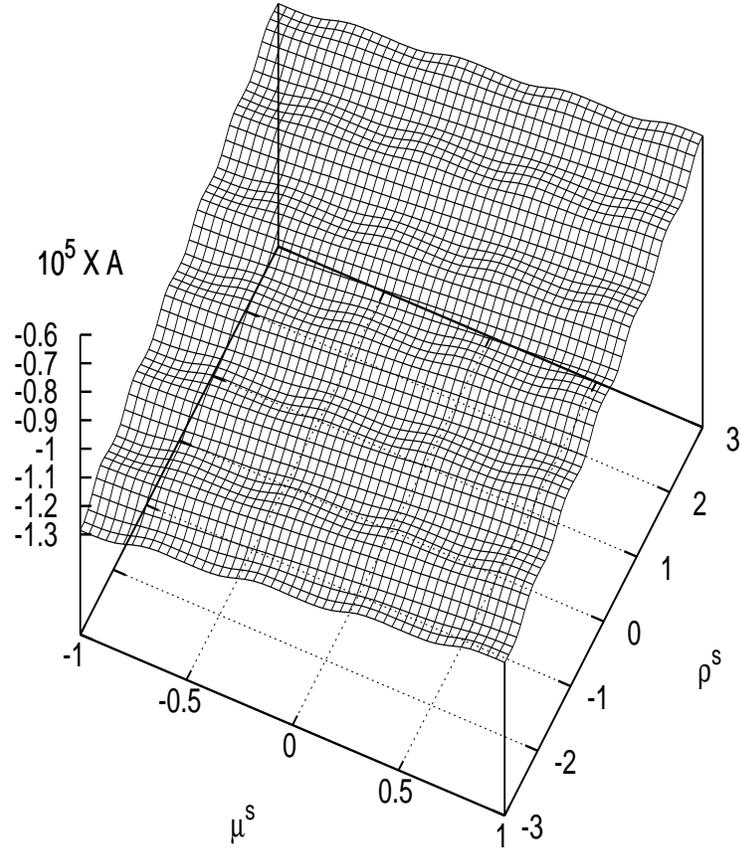} 
\vskip -0.3cm
\caption {PV asymmetry for $^{12}$C at $q$ = 500 MeV/c, 
$\omega$ = 120 MeV, and $\vartheta$ = 30$^{\mathrm o}$ as a function 
of $\rho^{\mathrm s}$ and $\mu^{\mathrm s}$.
}
\label{3d1}
\end{center}
\end{figure}

\begin{figure}[h]
\begin{center}
\includegraphics[height=14cm, width=12cm]{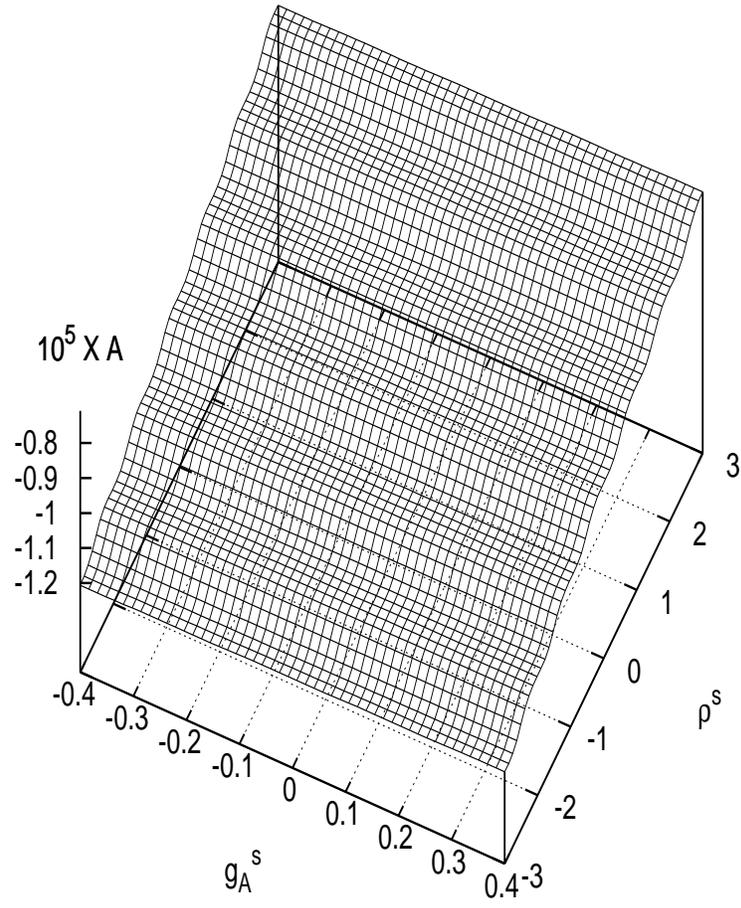} 
\vskip -0.3cm
\caption {The same in Fig. \ref{3d1}, but as a function of $\rho^{\mathrm s}$ and 
$g_{\mathrm A}^{\mathrm s}$}.
\label{3d2}
\end{center}
\end{figure}

\begin{figure}[h]
\begin{center}
\includegraphics[height=14cm, width=12cm]{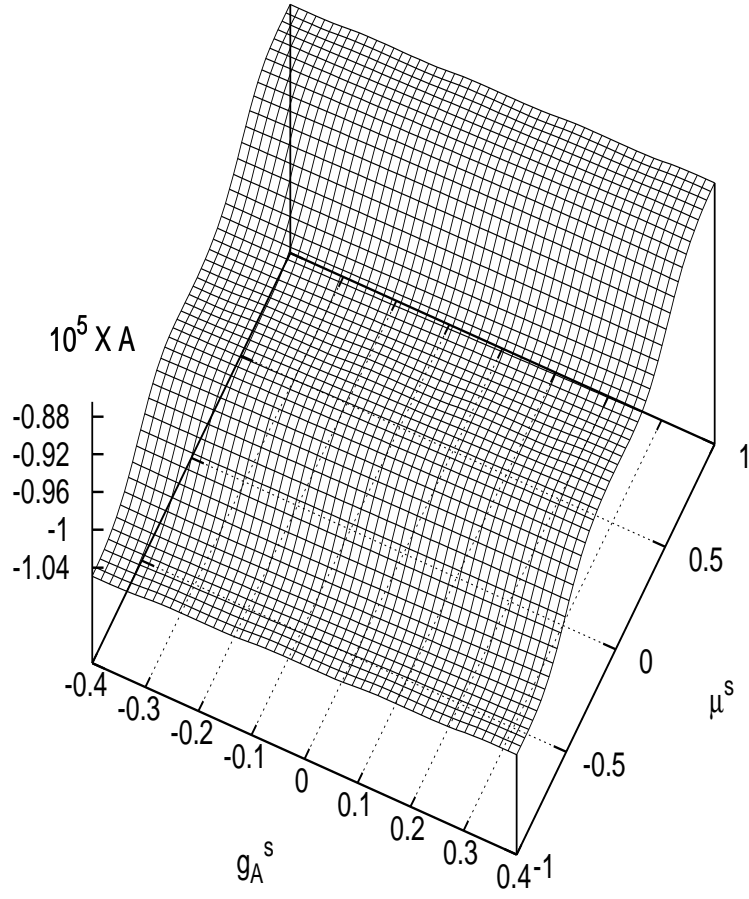} 
\vskip -0.3cm
\caption {The same in Fig. \ref{3d1}, but as a function of $\mu^{\mathrm s}$ and 
$g_{\mathrm A}^{\mathrm s}$}.
\label{3d3}
\end{center}
\end{figure}

The sensitivity of PV electron scattering to the effect of 
strange-quark contribution to the vector and axial-vector form factors, is 
shown in Figs. \ref{3d1}, \ref{3d2}, and \ref{3d3} for $^{12}$C at 
$q$ = 500 MeV/c, $\omega$ = 120 MeV, and 
$\vartheta$ = 30$^{\mathrm o}$ as a function of 
the strangeness parameters, 
$\rho^{\mathrm s}$, $\mu^{\mathrm s}$, and $g^{\mathrm s}_{\mathrm A}$.  
The range of their values is chosen according to Refs. \cite{beck1,aniol}. 
The asymmetry
reduces up to 40\% as $\rho^{\mathrm s}$ varies in the range  
$-$3 $\leq \rho^{\mathrm s}\leq $ $+$3, whereas it changes up to 15\% for
$-$1 $\leq \mu^{\mathrm s}\leq $ $+$1. We note that, according to HAPPEX results 
\cite{aniol},
$\rho^{\mathrm s}$ and $\mu^{\mathrm s}$ might have opposite sign, thus leading 
to a partial cancellation of the effects.
The sensitivity to $g^{\mathrm s}_{\mathrm A}$ is very weak.

\begin{figure}[h]
\begin{center}
\includegraphics[height=14cm, width=12cm]{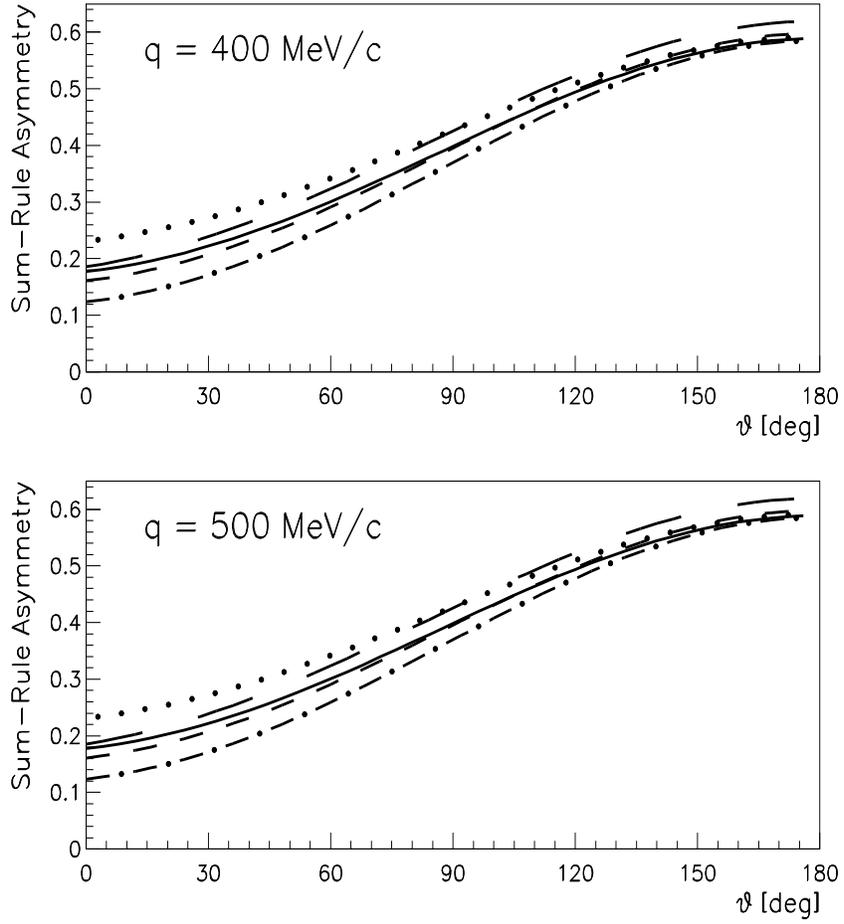} 
\vskip -0.3cm
\caption {The sum-rule asymmetry for $^{12}$C 
at $q$ = 400 and 500 MeV/c. 
Solid lines represent the results with no strangeness contribution,  
dashed (long-dashed) lines with $\mu^{\mathrm s}$ = $-$1 (+1), dotted 
(dot-dashed) lines with $\rho^{\mathrm s}$ = $-$3 (+3).
}
\label{asr1}
\end{center}
\end{figure}

In order to better show up the strangeness effects, in
Refs. \cite{alb1,donnelly,alb2,barbaro,amaro} they were
studied through the integrated sum-rule asymmetry: 
\inieq
\mathcal{ASR} = \frac {\int \diff \omega \left(
v^{}_{\mathrm{L}}R_{\mathrm{L}}^{\mathrm{AV}} + v^{}_{\mathrm{T}}
R_{\mathrm{T}}^{\mathrm{AV}} 
+v_{\mathrm{T}}'R_{\mathrm{T}}^{\mathrm{VA}}\right)/\tilde{X}_{\mathrm{T}}}
{\int \diff \omega \left(
 v^{}_{\mathrm{L}}R_{\mathrm{L}} + v^{}_{\mathrm{T}}R_{\mathrm{T}}\right)/
 X_{\mathrm{T}}} \ , \label{eq.asr}
\fineq
where the functions $\tilde{X}_{\mathrm{T}}$ and $X_{\mathrm{T}}$ are defined in
Refs. \cite{alb,barbaro}.
In Fig. \ref{asr1} the effect of the strange contribution
on the sum-rule asymmetry is 
shown for the scattering on $^{12}$C at $q$ = 400 and 500 MeV/c.  At
forward scattering angle the asymmetry is mainly dependent on the electric 
strangeness parameter $\rho^{\mathrm s}$, whereas the magnetic strangeness
parameter $\mu^{\mathrm s}$ becomes more important at backward scattering angle.
The sensitivity to the strange component of the axial form 
factor is weaker and only gives, at backward scattering angles, a modest effect 
that is not shown in the figure.
Similar results are obtained when different target
nuclei such as $^{16}$O and $^{40}$Ca are considered.

%%%%%%%%%%%%%%%%%%%%%%%%%%%%%%%%%%%%%%%%%%%%%%%
\section{Summary and conclusions}
\label{conc}

A relativistic approach to parity-violating quasielastic electron scattering, 
based on the spectral representation of the single-particle 
Green's function in terms of the eigenfunctions of the complex optical 
potential and of its Hermitian conjugate, has been presented. This approach has 
proved to be rather 
successful in describing inclusive electron scattering and charged-current 
neutrino-induced reactions. The effects of final state interactions are included
in a simple way that keeps flux conservation by using an optical potential  
consistently with exclusive processes. The imaginary part of the potential
accounts for the redistribution of the strength among different channels,
without any flux absorption.

%The method is applied within a relativistic framework to electromagnetic and 
%weak current 
%reactions for a momentum transfer up to 500 MeV/c.
The transition matrix elements are calculated using a single-particle model
obtained in the framework of the relativistic mean field theory for the
structure of the nucleus and applying the direct Pauli reduction for the 
scattering state. 

Calculations of the parity-violating response functions and asymmetry have 
been presented for $^{12}$C, $^{16}$O, and $^{40}$Ca target nuclei and 
for momentum transfers up to 600 MeV/c. The results of different 
approximations of final state interactions have been compared.
The effect of the optical potential and of the conservation of flux on the 
response functions is large.  Smaller effects are found on the asymmetry.
The sensitivity to the strange-quark content of the vector and axial-vector 
form factors has been investigated with different values of the parameters.
Forward-angle scattering may help to determine the electric strangeness whereas
backward-angle scattering may add more information about the magnetic 
strangeness
form factor.

%\begin{ack}

%\end{ack}

%%%%%%%%%%%%%%%%%%%%%%%%%%%%%%%%%%%%%%%%%%%%%%%

%\vskip 4cm

%%%%%%%%%%%%%%%%%%%%%%%%%%%%%%%%%%%%%%%%%%%%%%%%%% 
%%%%%%%%%%%%%%%%%%%%%%%%%%%%%%%%%%%%%%%%%%%%%%%%%%%
%%%%%%%%%%%%%%%%%%%%%%%%%%%%%%%%%%%%%%%%%%%%%%%%%%%
%%%%%%%%%%%%%%%%%%%%%%%%%%%%%%%%%%%%%%%%%%%%%%%%%%
%%%%%%%%%%%%%%%%%%%%%%%%%%%%%%%%%%%%%%%%%%%%%%%%%%%
%%%%%%%%%%%%%%%%%%%%%%%%%%%%%%%%%%%%%%%%%%%%%%%%%%%

%%%%%%%%%%%%%%%%%%%%%%%%%%%%%%%%%%%%%%%%%%%%%%%
%%%%%%%%%%%%%%%%%%%%%%%%%%%%%%%%%%%%%%%%%%%%%%%%%%%
%%%%%%%%%%%%%%%%%%%%%%%%%%%%%%%%%%%%%%%%%%%%%%%
%%%%%%%%%%%%%%%%%%%%%%%%%%%%%%%%%%%%%%%%%%%%%%%
%%%%%%%%%%%%%%%%%%%%%%%%%%%%%%%%%%%%%%%%%%%%%%%
%%%%%%%%%%%%%%%%%%%%%%%%%%%%%%%%%%%%%%%%%%%%%%%

%%%%%%%%%%%%%%%%%%%%%%%%%%%%%%%%%%%%%%%%%%%%%%%

%%%%%%%%%%%%%%%%%%%%%%%%%%%%%%%%%%%%%%%%%%%%%%%
%%%%%%%%%%%%%%%%%%%%%%%%%%%%%%%%%%%%%%%%%%%%%%%
%%%%%%%%%%%%%%%%%%%%%%%%%%%%%%%%%%%%%%%%%%%%%%%

\end{document}